# Considering Fundamental Rights in the European Standardisation of Artificial Intelligence: Nonsense or Strategic Alliance?

*Marion Ho-Dac*[1]

Professor of Private Law at Artois University,
Member of the Research Centre "Law Ethics and Procedures" (CDEP UR 2471), France
& Fellow at the Academy of International Affairs NRW, Germany
marion.hodac@univ-artois.fr

*Abstract:* In the European context, both the EU AI Act proposal and the draft Standardisation Request on safe and trustworthy AI link standardisation to fundamental rights. However, these texts do not provide any guidelines that specify and detail the relationship between AI standards and fundamental rights, its meaning or implication. This chapter aims to clarify this critical regulatory blind spot. The main issue tackled is whether the adoption of AI harmonised standards, based on the future AI Act, should take into account fundamental rights. In our view, the response is yes. The high risks posed by certain AI systems relate in particular to infringements of fundamental rights. Therefore, mitigating such risks involves fundamental rights considerations and this is what future harmonised standards should reflect. At the same time, valid criticisms of the European standardisation process have to be addressed. Finally, the practical incorporation of fundamental rights considerations in the ongoing European standardisation of AI systems is discussed.

## 1  Introduction

### 1.1  The EU AI Act as legal context

The Artificial Intelligence Act proposal (hereinafter "AI Act") was released in Spring 2021 by the European Commission[2] and is currently being negotiated within the European Union (hereinafter "EU") legislative process.[3] It sets up a uniform legal framework designed to ensure the free movement of AI systems in the EU internal market. It lays down a set of requirements and obligations for AI systems providers (and other stakeholders of the AI systems value chain) following a risk-based approach.[4]

---

[1] The author is a volunteer expert member of the French Standardisation Commission on AI, AFNOR, France. The opinions she expresses are her own.

[2] Proposal for a Regulation of the European parliament and of the Council laying down harmonised rules on artificial intelligence (Artificial Intelligence Act) and amending certain union legislative acts, COM/2021/206 final.

[3] Procedure 2021/0106/COD available online at:
https://eur-lex.europa.eu/legal-content/EN/HIS/?uri=celex:52021PC0206

[4] See the short presentation on this approach by the European Commission, available at: https://digital-strategy.ec.europa.eu/en/policies/regulatory-framework-ai. Cf. Mahler, Tobias, "Between risk management

The AI Act gives a major role to both the protection of fundamental rights, generally understood as rights inherent to all human beings on an equal basis (e.g. the right to life and liberty, the freedom of opinion and expression)[5] and to standardisation, i.e. the adoption by consensus by a recognised body of rules, guidelines or characteristics in a given sector.[6] On the one hand, the Act proposal provides that the high-risk AI systems shall comply with certain mandatory requirements in line with the European public interests of health, safety and the defence of fundamental rights. On the other hand, it lays out that those requirements on (inter alia) risk management, data governance, transparency, human oversight may be translated into harmonised standards, developed within the European standardisation bodies.

Reflecting on whether and how fundamental rights considerations should be taken into account, or even integrated, in the making of European AI standardisation is essential for all stakeholders – the industry, SMEs, researchers and representatives of civil society. First, fundamental-rights-oriented standards are not self-evident. Standardisation is rather a domain reserved for the transcription of the technical state of the art in a given field than the place where human rights are analysed or implemented. Second, the direct involvement of all stakeholders of the AI ecosystem in the "standardisation effort", including potential fundamental rights considerations, is highly desirable to ensure sufficient legitimacy and adequacy of future standards.

## 1.2 The inclusion of the AI Act in the New Legislative Framework

When European standardisation was first introduced in the mid-1980s, the aim was to limit the harmonisation of the national laws of the European Community Member States – which had become too extensive – to "essential requirements" of safety and health or reflecting other public interests.[7] The rest of the market regulation was left to technical manufacturing specifications translated into standards. European standardisation organisations (hereinafter referred to as "ESOs")[8] have been set up to develop these standards. The introduction of the New Legislative Framework (hereinafter "NLF") in 2008 marked a new regulatory step for the internal market for goods, completing the legal acquis, particularly in the field of conformity assessment and market surveillance.[9] The objective was to enhance the effectiveness of the regulatory framework by strengthening the monitoring – *ex ante* by notified bodies and *ex post* by public surveillance authorities – of compliance with EU law by economic operators. At the same time, the EU Regulation on standardisation allows the European Commission (hereinafter "the Commission") to ask the ESOs to draw up "harmonised standards". They are "voluntary technical or quality

---

and proportionality: The risk-based approach in the EU's Artificial Intelligence Act Proposal" (September 30, 2021). *Nordic Yearbook of Law and Informatics*, Available at SSRN: https://ssrn.com/abstract=4001444

[5] As a global reference, see the Universal Declaration of Human Rights, 10 December 1948, United Nations.

[6] See K. Jakobs, ICT Standardization, *Encyclopedia of Information Science and Technology*, 4th Edition, 2018, p. 13.

[7] Council Resolution 85/C 136/01 of 7 May 1985 on a new approach to technical harmonisation and standards.

[8] The ESOs are the European Committee for Standardisation (CEN), the European Committee for Electrotechnical Standardisation (CENELEC) and the European Telecommunications Standards Institute (ETSI).

[9] Regulation (EC) No 765/2008 of the European Parliament and of the Council of 9 July 2008 setting out the requirements for accreditation and market surveillance relating to the marketing of products and Decision No 768/2008/EC on a common framework for the marketing of products. Then Regulation (EU) 2019/1020 of 20 June 2019 on market surveillance and product conformity, OJ L 169, 25 June 2019, p. 1-44 (amending Regulation 765/2008).

requirements with which products, manufacturing processes or services [...] may comply"[10], with the specificity that they are "adopted on the basis of a request made by the Commission for the application of Union harmonisation legislation"[11] – as it will be the case regarding the AI Act –.

The Commission, but also all stakeholders, must therefore ensure that these standards perfectly match the legal requirements laid out in EU law.[12] However, there is a certain mistrust of harmonised standards and their adoption process by some members of the civil society, including academics, which is explicitly confirmed in the discussions on the AI Act.[13] Standards are private, non-binding rules. Consequently, on may argue that it is questionable and even critical that they can, at the same time, "form part of EU law"[14] and thus have law-like value.[15] Indeed, they are not subject to democratic debate for their adoption and their content is not published *in extenso* in the Official Journal of the EU (hereinafter "OJEU") as is EU law.[16]

Finally, the AI Act proposal fits into the context of the NLF. It contains its above-mentioned main regulatory features: harmonisation of national laws limited to essential requirements, obligation for providers to carry out conformity assessment, establishment of European market surveillance based on a network of national authorities and reference to harmonised standards. What is unusual, however, is the recurrent reference made by the AI Act to fundamental rights.

## 1.3     The unusual reference to fundamental rights

Beyond the traditional "internal market" approach, the AI Act contains an original feature that should be noted from the standpoint of standardisation. The public interests that the Act seeks to protect cover an innovative triptych: health, safety and *the protection of*

---

[10] Recital 1 and Article 2(1) Regulation (EU) No 1025/2012 of the European Parliament and of the Council of 25 October 2012 on European standardisation, OJ L 316, 14 November 2012, p. 12-33.

[11] Article 2(1)(c), *op. cit.* A complementary specificity of European harmonised standards lies in the fact that "the Commission shall [if certain conditions are fulfilled] publish a reference of such harmonised standard without delay in the *Official Journal of the European Union* or by other means in accordance with the conditions laid down in the corresponding act of Union harmonisation legislation", Article 10(6), *op. cit.* See also below §1.3 of this contribution.

[12] In this sense, the EU Regulation on standardisation (*op. cit.*) lays down procedures for the assessment of such standards and the possibility for Member States and the European Parliament to object to them if they do not fully meet the requirements of secondary legislation (Article 11). In addition, it is worth mentioning that the Court of Justice of the European Union (CJEU) ruled on the "justiciability" of harmonised standards, which it considers to be a matter of Union law. See CJEU, 27 October 2016, C-613/14, *James Eliott Construction Limited*, Case C-613/14, EU:C:2016:821, esp. point 43. In the context of the procedure of formal objections to a harmonised standard, the General Court and the Court of Justice can hear actions for annulment against such a standard. See for instance Cases C-475/19 P and C-688/19 P.

[13] See M. Ebers, "Standardizing AI – The Case of the European Commission's Proposal for an Artificial Intelligence Act", *DiMatteo, L., Poncibò, C., & Cannarsa, M. (Eds.). (2022), The Cambridge Handbook of Artificial Intelligence: Global Perspectives on Law and Ethics, Cambridge University Press;* M. Veale & F. Zuiderveen Borgesius, "Demystifying the Draft EU Artificial Intelligence Act", *Computer Law Review International*, 4/2021, p. 97-112, esp. §54 et seq.

[14] *James Elliott*, Case C-613/14, *op. cit.*, point 40.

[15] For the record, the implementation of European harmonised standards ("hEN") whose reference is published in the Official Journal of the EU, by organisations for their products – such as AI systems – constitute a presumption of compliance with the essential requirements laid down in EU law – such as the said requirements of the AI Act –. On the drafting procedure before CEN see: https://boss.cen.eu/developingdeliverables/pages/en/pages/enforojeu/

[16] *Ibid*.

*fundamental rights*. While the first two are classic, the reference to fundamental rights is unprecedented in NLF secondary legislation. It is the Court of Justice of the EU that has integrated the protection of fundamental rights, on a case-by-case basis, into the law of economic freedoms of movement through its case law.[17] In EU primary law, reference should be made to the Charter of Fundamental Rights of the European Union (hereinafter "EU Charter") which is part of primary EU law.[18] More broadly, the EU is based, from a constitutional standpoint, on core values – the Union values – listed into Article 2 of the EU Treaty and which includes inter alia "the values of respect for human dignity, freedom, democracy, equality, the rule of law and *respect for human rights*".[19]

This peculiar fundamental rights' dimension of the AI Act is justified by the "specific characteristics (e.g., opacity, complexity, dependency on data, autonomous behaviour) [of AI systems that] can adversely affect a number of fundamental rights".[20] As a response, the Act sets up a risk-based approach, including fundamental rights considerations. Interestingly both the internal market policy competence (Article 114 TFEU) and the data protection competence (Article 16 TFEU) are mentioned as legal basis of the Act. This could explain the mixed approach – market- and fundamental rights-oriented – followed by the text.[21]

As harmonised standards are supposed to reflect the said public interests, awareness and expertise in the field of EU fundamental rights should thus be necessary among standardisation actors. The crucial question is therefore whether ESOs will deal with the concrete scrutiny, or even inclusion of fundamental rights considerations into AI standards and, if yes, how they will proceed concretely.

It is necessary to study how the (draft) AI Act and the (draft) Standardisation Request to the ESOs in support of safe and trustworthy AI handle this complex and sensitive relationship between fundamental rights and AI regulation. The issue is openly controversial. For some actors, such a relationship is neither necessary nor relevant: it is not the role of standards to deal with fundamental rights. It belongs to the EU as political entity based on its constitutional framework. For others, on the contrary, it is essential to link fundamental rights to AI standards. The interplay between fundamental rights and AI systems should be considered in the standards-setting process, and even the translation of those rights by design into AI systems could be envisaged. This is an important way of addressing individual and societal needs, as some recent international standards do, via fundamental rights considerations and features.

### 1.4 The main lines of the chapter

Against this background, this chapter aims to show that the draft texts of the AI Act and of the Standardisation Request clearly link standardisation to fundamental rights. However, these texts do not provide any guidelines that specify and detail this relationship and its operationalisation. This is regrettable. Hence, it is necessary to invite all AI stakeholders, including EU institutions and the Member States, to concretely address the issue of so-called "trustworthy" AI based inter alia on fundamental rights considerations. The chapter demonstrates that this could be achieved, at least partly, through standardisation. This should not be understood as a withdrawal or decommitment of the EU legislator in defending the fundamental rights of European citizens. Harmonised standards aim at supporting EU law. Therefore, they can usefully raise awareness and provide understanding of fundamental rights implication in AI context, and thus fundamental rights compliance of AI systems, among AI practitioners and the tech industry.

---

[17] See for instance on human dignity in the field of freedom to provide services, ECJ, 14 October 2004, *Omega Spielhallen*, Case C-36/02, EU:C:2004:614.

[18] OJ C 236, 26 October 2012, p. 391-407. See also Article 6 of the EU Treaty.

[19] *The emphasis is mine*.

[20] Point 3.5., *Explanatory Memorandum*, AI Act proposal, *op. cit.*

[21] By comparison, the situation is different in the General Data Protection Regulation which is based on Article 16 TFEU and not on Article 114 TFEU. Although the latter provision belongs to the internal market policy, the former provides for the protection of personal data as a fundamental right (as recalled in the first sentence of Recital 1 of the GDPR). See Regulation (EU) 2016/679 of the European Parliament and of the Council of 27 April 2016 on the protection of natural persons with regard to the processing of personal data, OJ L 119, 4 May 2016, p. 1-88.

## 2    The place of fundamental rights and standards in the AI Act

### 2.1    The twofold reference to fundamental rights and standards in the wording of the AI Act

#### 2.1.1  *Fundamental rights in the wording of the AI Act*

In the AI Act, fundamental rights derived from the EU Charter have a very important place regarding their domain of intervention, i.e., the internal market. They are systematically mentioned as "public interests" to be protected alongside health and safety. Their guarantee, in the context of the design, development and use of AI, is presented as justifying the establishment of a uniform mandatory framework for high-risk AI systems.[22] In more details, the AI Act specifies that the measure of the negative impact that an AI system may have on the fundamental rights guaranteed by the EU Charter is an essential criterion for implementing the risk pyramid adopted by the Act and ranking AI systems within.[23] The higher the risk of infringement of a fundamental right, the more the AI system in question should be placed at the top of the pyramid. In that respect, AI systems that present unacceptable risks for individuals and the society (e.g., in the field of social scoring and biometrics) shall be subject to an *a priori* ban, justified by their *de jure* violation of European fundamental rights.[24] Regarding AI systems considered as high-risk, they shall be subject to strict supervision (as already explained). This fundamental rights risk-based approach is also the methodology that should be used by the Commission when adapting the list of high-risk systems and adding new systems[25].

According to the AI Act, the providers of high-risk AI systems will (inter alia) have to put in place a risk management system to address the potential risks[26] – described as any "significant adverse impact"[27] – to "health, safety and fundamental rights".[28] Similarly, in the context of market surveillance, national authorities should be able to suspend or prohibit the marketing of an AI system that does not comply with the requirements of the

---

[22] Recital 13, AI Act proposal, *op. cit.*

[23] Recital 28, *op. cit.*

[24] Article 5, *op. cit.*

[25] Article 7, *op. cit.*

[26] Article 9, *op. cit.*

[27] Recital 27, *op. cit.*

[28] Article 9, EU Council General Approach, 6 December 2022, ST 15698 2022 INIT.

Act and/or presents risks to health, safety or fundamental rights[29] regardless of the system's classification on the risk pyramid.[30]

### 2.1.2 Standards in the wording of the AI Act

In the explanatory memorandum of the AI Act, technical standards are presented as the means to implement "concretely" the common mandatory requirements[31] applicable to high-risk AI systems, from their design and throughout their life cycle.[32] Standardisation is therefore an optional and tailor-made path for operators to ensure the most effective compliance with the Act.[33]

The AI Act builds on the EU Regulation on standardisation to define, in Article 40, the legal contours that will be given to harmonised standards to be drafted by ESOs. Those standards should technically and comprehensively translate the requirements of Articles 9 to 15, in Chapter II, of the AI Act and their references be published in the OJEU, after approval of the European Commission. Then, high-risk AI systems that comply with the harmonised standards (in whole or in part) will be presumed to comply with Chapter II of the Act. These harmonised standards will have to specify and explain their exact coverage of the said requirements (in terms of risk management, data quality, automatic recording of events, transparency, human oversight or robustness, etc.), so that the scope of the presumption of conformity is clearly established.

Against this background, the quality management system that providers of AI systems will have to put in place in order to ensure their compliance with the AI Act refers to the use of "technical specifications" and therefore, most likely, the implementation of harmonised standards.[34] Conformity assessment,[35] another mandatory step for AI providers under the Act, understood as "the process of verifying whether the [essential] requirements […] relating to an [high-risk] AI system have been fulfilled",[36] is also relying on standards.[37]

The importance of European standards as well as fundamental rights, each with their own meaning and scope, is clear from the AI Act. The question to be addressed now is their interrelation. How does the EU intend to link these two normative tools which a priori belong to very different conceptual spheres?

---

[29] Articles 65, 67 et 68, AI Act proposal, *op. cit.*

[30] Article 65, *op cit.*

[31] See Chapter 2, Title III (Articles 8 to 15) of the AI Act proposal, *op cit.*

[32] Cf. The 'Blue Guide' on the implementation of EU products rules 2022, 2022/C 247/01, OJ C 247, 29 June 2022, p. 1-152, esp. §2.1 et §5.2.3.

[33] Standards are one technical solution chosen by a provider to comply with the requirements laid down by EU law, among others, such as technical specifications that may be developed in accordance with general engineering or scientific knowledge.

[34] Article 17 (e), AI Act proposal. If the provider decides not to rely on the standards, it is asked to explain "the means to be used to ensure that the high-risk AI system meets the requirements set out in Chapter 2 [of the AI Act]".

[35] Articles 19 and 43, *op. cit.*

[36] Article 3(20), *op. cit.*

[37] In particular, the AI Act proposal requires Notified Bodies to participate in the activities of ESOs or, at least, to keep themselves informed of the applicable standards and their status (Article 33, §11, AI Act proposal).

## 2.2 The link between fundamental rights and AI standards

### 2.2.1 AI standards' objective to support the respect of fundamental rights

The General Approach of the EU Council on the AI Act makes the process of adoption of harmonised standards explicit.[38] It sets out a series of objectives that future standards should meet under the supervision of the European Commission. These objectives are essential, in our view, as they should have an influence on the content of future harmonised standards. Alongside the promotion of innovation, the enhancement of multi-stakeholder governance and the strengthening of international cooperation in the field of standardisation,[39] one key objective focuses on the protection of European values. Among these values, as already mentioned above, Article 2 of the Treaty on EU lists core human rights, such as the respect for human dignity and non-discrimination, crucial political principles of democracy and the rule of law, as well as, more generally, the respect for fundamental rights. It will be up to the ESOs to ensure that this objective is met.[40] This does not mean that standards will have *per se* to implement or balance fundamental rights in AI context but rather that their content should reflect fundamental rights considerations. Therefore, standards implementation by organisations should make it possible to limit the risks of their AI systems, starting with the risks of fundamental rights' violation.

The draft Standardisation Request, released by the Commission in December 2022, largely takes up the objectives specified in Article 40 of the General Approach. It similarly includes the need for standards to play a role in ensuring the protection of fundamental rights in AI context.[41] Following the rationale of the Regulation on standardisation, future standards should take into account EU policy objectives. In the field of AI, this implies that AI systems placed in the EU market "are used in compliance with fundamental rights".[42] Hence, it may suggest that the design and development of AI system should consider the question of fundamental rights or even – following a maximalist reading – that AI systems should comply with human rights by design.

This is coherent with the EU promotion of a human-centred approach of digital governance. In that respect, the Commission has recently put forward a 'Declaration on European Digital Rights and Principles'[43], putting people at the centre of digital transformation.[44] This policy vision is rooted in "digital humanism" which consists in making the well-being of humans in the digital ecosystem the priority.[45] This includes digital constitutionalism,[46] based on the primacy of human-being as a social individual and citizen who enjoys the protection of fundamental rights. The defence of fundamental rights must be equally guaranteed in the physical world and online. The use of AI is by nature immaterial and connected to the use of the internet or other connectivity capabilities. While it is primarily the responsibility of States and the EU to ensure this "equivalent protection", tech organisations must also be made accountable. This implies a multi-stakeholder regulation, including fundamental rights consideration, of the digital and AI ecosystem. This larger political context clearly underpins the ongoing AI systems regulation.

---

[38] Contrary to the Commission's AI Act proposal and based on a Standardisation Request of the Commission, itself based on Article 10 of the Regulation on standardisation, *op. cit.*

[39] Articles 40 (2), (b), (c) and (d), General Approach, *op. cit.*

[40] Article 40, *in fine, op. cit.*

[41] Recital 2, draft Standardisation Request, *op. cit.*

[42] Recital 13, *op. cit.*

[43] COM(2022)28 final, 26 January 2022.

[44] Recital 4, *op. cit.*

[45] Cf. J. Nida-Rümelin, N. Weidenfeld, *Digital Humanism, For a Humane Transformation of Democracy, Economy and Culture in the Digital Age*, Springer, 2022 Open Access. See the Digital Humanism Observatory in Vienna and its work. See also the summary of the High-level workshop on Digital Humanism and Artificial Intelligence, organised by the members of the Slavkov format on 3 March 2022, available here: https://data.consilium.europa.eu/doc/document/ST-9415-2022-INIT/en/pdf.

[46] See recently De Gregorio, Digital *Constitutionalism in Europe, Reframing Rights and Powers in the Algorithmic Society*, Cambridge Press, 2022.

This position is also reflected in the European Commission's list of future standardisation deliverables and leads to integrating fundamental rights considerations in standard-setting process and AI standards.

### 2.2.2 Fundamental rights' implications for European AI standards

Following the EU holistic approach, future standards on AI will have to reflect both the defence of European public interests, including the protection of fundamental rights, and safeguard flexibility, legal certainty and efficiency in favour of the growth and innovation of AI industry. This balance is reflected in the draft request for standardisation by the combination of a list of deliverables in Annex I and the associated requirements for each deliverable in support of the AI Act in Annex II. Ten items (for now) should give rise to normative deliverables with some key characteristics, pursuant to the future regulation: (1) risk management, (2) governance and quality of data sets, (3) recording and automatic logs, (4) transparency and information to users, (5) human oversight, (6) accuracy, (7) robustness, (8) cybersecurity, (9) quality management system and (10) conformity assessment.

Regarding the requirements applicable to all future standards, they "shall reflect the generally acknowledged state of the art in order to minimise the risks to health and safety and fundamental rights of persons as guaranteed in the Charter of Fundamental Rights of the EU which arise from the design and development of AI systems in view of their intended purpose".[47] In other words, the technical specifications that will be developed (e.g. characteristics, indicators, tests, metrics, guidelines, etc.) aim to manage – and ultimately mitigate – the risks of AI systems, including risks to the fundamental rights of individuals. In this context, the drafting of standards cannot be entirely technical, based on computer science and engineering knowledge. It must have a societal dimension that can be linked to humanities and social sciences considerations and knowledge, including law. This does not mean that standards are becoming "public" tools for implementing fundamental rights and extending hard law. This would overstate their original nature, which is both private and mainly technical. Plus, from a legal point of view, (binding) provisions on fundamental rights requirements applicable to AI systems, under the AI Act and other EU law instruments, are in theory sufficient to bring an action against an AI provider who has failed to comply with them and to sanction him. One main reason why standards need to take fundamental rights into account is to provide AI providers with techniques and methodologies to analyse, anticipate and adapt the effects that the AI systems they design, develop and/or deploy may have on the rights of individuals. This could imply inter alia that AI stakeholders have a basic understanding of "fundamental rights" as a discipline: Which rights are at stake in different AI use cases? What are their meaning and characteristics? What are the conditions for enjoying those rights and what are their limits if any? etc. This echoes fundamental rights impact assessment already existing in advanced technologies domain.[48]

---

[47] Annex II (1) §1, draft Standardisation Request, *op. cit.*

In that respect, the draft Standardisation Request provides that "CEN and CENELEC shall ensure the appropriate involvement in the standardisation work of EU small and medium enterprises, civil society organisations, and gather *relevant expertise in the area of fundamental rights*".[49] This is certainly not usual in the European standardisation ecosystem. It will be very challenging in practice for the ESOs within the Joint-technical Committee on AI ("JTC 21").[50]

In this context, it is central to address the issue of integrating fundamental rights considerations in AI standardisation work at the European level.

## 3 The inclusion of fundamental rights considerations into European AI standards

### 3.1 The questionable legitimacy of standards to ensure respect for fundamental rights by AI systems

#### 3.1.1 The lack of a constitutional dimension to standardisation

Given its importance in EU internal market law, it is understandable that standardisation may be the subject of concern or even criticism.[51] It is even more critical when fundamental rights considerations are at stake, as the latter are the highest embodiment of hard law. This is mostly the case from a legal perspective. European standardisation is pointed out as lacking "constitutional legitimacy". This reading is based on the notion of Constitution as a legal norm qualified as fundamental, safeguarding human freedom based on a list of human rights.[52] The existence of such a Constitution in a given legal system is considered a condition for democracy.[53] At the same time, the Constitution ensures democracy.[54] In the EU, the respect of fundamental rights enshrined in its "Constitution" – i.e. the European treaties including the EU Charter – is at the very heart of the EU democratic nature.[55] And it refers more broadly to the rule of law, as core European value recently put under the spotlight, inter alia, in the context of Polish and Hungarian national reforms deemed contrary to this European value.[56]

---

[48] For instance, the Government of the Netherlands has developed a "Fundamental Rights and Algorithm Impact Assessment, named (FRAIA) [that] helps to map the risks to human rights in the use of algorithms and to take measures to address these risks", March 2022, online resource. See also *infra*, Section 3 of this chapter.

[49] *Ibid. This emphasis is mine.* See also Recital 14 and Article 2, draft Standardisation Request, *op. cit.* See the reference to "given the fundamental rights implications of the European standards and European standardisation deliverables requested […]".

[50] On CEN-CENELEC JTC 21, see *infra*, Section 3.2.

[51] See M. Eliantonio and C. Cauffman (Eds.), *The Legitimacy of Standardisation as a Regulatory Technique in the EU*, Elgar, 2020; M. Lanord-Farinelli, « La norme technique: une source du droit légitime? », *Revue française de droit administratif*, 2005, p. 738 et seq.

[52] See P. Brunet, "Constitution", *Encyclopædia Universalis*, online version, 3 Mars 2023.

[53] *Ibid.*

[54] According to D. Gaxie, "The specificity of a democratic system is that the people governed are supposed to be at the same time governors, associated with the main decisions affecting the everyday life. And it is because the people are both subjects (i.e., subject to political power) and sovereigns (holders of this power) that democratic systems are supposed to act in the interests of the people", *in* "Démocratie", *Encyclopædia Universalis*, online version, 3 Mars 2023 (*our translation*).

[55] See K. Lenaerts, P. Van Nuffel and T. Corthaut, *EU Constitutional Law*, Oxford University, 2021.

What does this EU constitutional dimension imply as regards standardisation? There are at least two main critical implications: one is internal to the standards-making and the other is external to it, lying in EU institutional sphere.

Standards are non-State rules, since they are adopted by consensus within the ESOs, which are themselves private non-profit organisations. Their drafting process is not as open and transparent as a legislative process in a democracy. It is not subject to any parliamentary debate but limited to the adoption of a consensus by the stakeholders of each draft standard. And, in practice, the said stakeholders are dominated by the leading international economic operators, mostly affected by the given standard in preparation. In that respect, the whole society is not represented. When they are released, standards are not freely and publicly accessible to citizens or civil society representatives.[57] They are subject to intellectual property rights.[58] Their access is therefore, in principle, subject to a fee. It is also true for harmonised standards, even though their implementation allows operators to benefit from a presumption of compliance with EU law. Yet, as already mentioned previously, only their reference is published in the OJEU.[59] This makes critical analysis of standards by academics more limited. It could also severely undermine, at least indirectly, the access to justice and legal remedies. Evidence of a violation of mandatory rules of EU law, such as the future AI Act, for the benefit of alleged victims, would benefit from knowledge of the standards that detail the practical implementation of these rules. The fact that they will not be published in full makes their access more difficult, but more fundamentally reduce their "visibility" for legal practitioners. The latter will not easily think to consult them for a case, although they may constitute a precious source of information.

Another set of criticisms, which is also based on the European constitutional dimension, is directed at the EU institutions. The very fact that standardisation is interconnected with fundamental rights raises important questions. As already noted above, this is not the traditional function of standards that aim to reflect the technical state of the art in a given field. Beyond those aspects, the societal impact of standards may be a factor in their making. But this is less far-reaching than the integration of fundamental rights considerations into the standard itself as it emerges from the AI Act and the Standardisation Request. This raises at least two important problems.

---

[56] See recent CJEU judgments in Cases C-156/21 *Hungary v Parliament and Council* and C-157/21 *Poland v Parliament and Council*. See also L. Pech & K. Scheppele (2017), "Illiberalism Within: Rule of Law Backsliding in the EU", *Cambridge Yearbook of European Legal Studies*, 19, 3-47.

[57] For a dispute (application for annulment) on the refusal by the European Commission to grant (free) access to NGOs to harmonised standards adopted by CEN (in the field of toys safety) see the judgment of the General Court, 14 July 2022, *Public.resource.Org and Right to Know v Commission*, case T-185/19. The application was dismissed (see Appeal Case before the Court of Justice, C-588/21 P).

[58] Point 43 of the above-mentioned judgment, the General Court held that "the Commission demonstrated that disclosure of the requested harmonised standards could […] undermine the commercial interests of CEN or its national members and that the risk of those interests being undermined was reasonably foreseeable".

[59] Point 107 of the above-mentioned judgment, the General Court rules that "the applicants do not state the exact source of a 'constitutional principle' which would require access that is freely available and free of charge to harmonised standards, they do not in any way explain the reason why those standards should be subject to the requirement of publication and accessibility attached to a 'law', inasmuch as those standards are not mandatory, they produce the legal effects attached to them solely with regard to the persons concerned, and they may be consulted for free in certain libraries in the Member States".

On the one hand, one may have the impression that the Union and the European legislator are delegating the making of hard law (i.e., fundamental rights implementation) to private bodies. This criticism of normative delegation is important and has already been scrutinised by legal scholars.[60] It echoes, more broadly, the democratic deficit that surrounds the adoption of European standards. Standardisation bodies have neither the political and legal competence nor the expertise to regulate fundamental rights. Ironically, private organisations are certainly the first to point this out (probably to escape any fundamental rights considerations).

On the other hand, certain standards are given a strong scope in the European normative context, almost equivalent to the law. This is of course the case of harmonised standards, but not only.[61] In technical domains, EU secondary law refers in some cases to international standards to specify the way of implementing certain legal provisions.[62] In this context, the Court of Justice of the European Union (CJUE) ruled, on one side, that "technical standards determined by a standards body, such as ISO, and made mandatory by a legislative act of the European Union are binding on the public generally only if they themselves have been published in the *Official Journal of the European Union*".[63] But, on the other side, the CJEU hold that "where undertakings have access to the official and authentic version of the standards referred to in […] [the EU secondary law instrument concerned], those standards and, therefore, the reference made thereto by that provision are binding on them"[64]. In the field of personal data protection, the General Data Protection Regulation (GDPR) has provided for a data protection impact assessment (DPIA) scheme. Harmonised standards were requested for in the context of the GDPR adoption process[65] aiming, inter alia, to address and manage privacy and personal data protection issues by design.[66] Based on the international standard ISO/IEC 27701 on Privacy Information Management System (PIMS) refined for the European context, a standard on data protection and privacy by design and by default was eventually adopted within the ESOs (i.e., EN 17529:2022).[67] In practice, some national data protection regulators in the EU Member States encourage organisations to refer to these standards for the purpose of legal compliance.[68] Based on this "law-like dimension" of standards, the question of their non-constitutional making-process and subsequent closed access should be addressed urgently. This also shows, in our view, that when the regulatory framework appeals to fundamental rights (as it is in the GDPR or in the AI Act proposal), it is not always sufficient in itself, at least from an operational perspective. Respect of fundamental rights is a highly complex and case-based issue, whereas the recipients of the legal provisions need legal predictability. In order to anticipate this case-by-case dimension of the law dealing with fundamental rights considerations, recourse to standardisation may be seen as a complementary approach.

---

[60] See C. H. Hofmann, "The Integration of Global Standards into the EU as 'regulatory Union'", 7 October 2022, University of Luxembourg Law Research Paper No. 2022-006; H-W. Micklitz, R. van Gestel, "European integration through standardization: How judicial review is breaking down the club house of private standardization bodies", (2013), 50, *Common Market Law Review*, Issue 1, pp. 145-181.

[61] For a typology of the "entry point for standards into EU law" (based on a broader concept of standard), see C. H. Hofmann, "The Integration of Global Standards into the EU as 'regulatory Union'", *op. cit.*, p. 8 ff.

[62] For instance, see the reference to ISO standards for the measure of the tar, nicotine and carbon monoxide yields in cigarettes, *in* Directive 2014/40/EU of the European Parliament and of the Council of 3 April 2014 on the approximation of the laws, regulations and administrative provisions of the Member States concerning the manufacture, presentation and sale of tobacco and related products and repealing Directive 2001/37/EC Text with EEA relevance, JO L 127 du 29 April 2014, p. 1–38, Article 4.

[63] CJEU, 22 February 2022, *Stichting Rookpreventie Jeugd and Others*, Case C-160/20, ECLI:EU:C:2022:101, point 53. In our opinion, publication of the standard *in extenso* should be distinguished from publication of the standard mere reference in the Official Journal, as is the case for European harmonised European. See *infra*, our final conclusions. For a comment on this CJEU case, see A. Volpato, "Transparency and legal certainty of the references to international standards in EU law: smoke signals from Luxembourg?", 17 March 2022, *Maastricht University Blog* (online).

[64] Point 52. In our view, there is a contradiction or at least an uncertainty between these two levels of analysis. Simple access, on private initiative, to a standard via a national standardisation organisation (NSO) in a Member State triggers its enforceability based on its reference in secondary legislation. Who then bears the burden of proof of access to the standard? And what does "access" concretely mean? Could an economic operator therefore be held liable for a breach of Union law for non-compliance with an ISO (or CEN-CENELEC) standard covered by a provision of EU secondary law from the moment it had access to that standard (even potentially, by being a member of a NSO, for example)? These questions could be particularly relevant in the context of AI, where ISO has a significant standardisation activity.

[65] Standardisation request in support of Directive 95/46/EC on personal data protection and of Union's security industrial policy, 20 January 2012, C(2015) 102 final.

[66] Annex 1 to Commission implementing decision, C(2015) 102 final, op. cit.

Against this background, we argue that those "constitutional" weaknesses have to be treated as such[69] and should not prevent from exploring fundamental rights considerations that are essential in AI standardisation.

### 3.1.2 The primacy of the fundamental rights approach

First, fundamental rights considerations in standardisation are necessary as they represent the highest level of public interests' implication. Based on the NLF, standards must serve public interest. If they were originally conceived on a technical level, in recent years they have also been considering their impact on individuals, society and the environment.[70] At the same time, the predominant influence of the industry in the standards-setting process had led to a greater influence of economic private interests at the expense of the public interest.[71] In this context, the European standardisation framework was reformed in 2012 to strengthen transparency and multi-stakeholder participation, in particular that of consumers, workers and civil society representatives.[72] Though, this was not sufficient.[73] A further reform of the ESOs governance was adopted in December 2022 and AI policy, in the context of the AI Act adoption has been chosen as a "test-case" to improve the standardisation system.[74] The Commission called on the ESOs "to modernise their governance to fully represent the *public interest* and interests of SMEs, civil society and users and to facilitate access to standards."[75] One of the main objectives of the reform has been to strengthen the role of national standardisation bodies in the adoption of standards. Article 10(2) of the EU Regulation on standardisation has been amended in order to ensure that the national standardisation bodies have exclusive voting rights for harmonised standards, thus ruling out the "direct" voting rights of private organisations. Even though this legislative change will only be applicable from July 2023, the draft Standardisation Request in the field of AI only refers to the European Committee for Standardisation (CEN) and the European Committee for Electrotechnical Standardisation (CENELEC) as competent ESOs, implicitly excluding the European Telecommunications Standards Institute (ETSI), whose current governance is at odds with the new text. This marks a certain shift in the weight of global industry in the European harmonised standards making-process. This will not be sufficient for the safeguard of public interests – including fundamental rights considerations – by future standards, but it already constitutes a first development in the right direction.

---

[67] With no citation in the OJEU expected for the GDPR. See here: https://standards.cencenelec.eu/dyn/www/f?p=205:110:0::::FSP_PROJECT:63633&cs=143EAD5D19379C232793068EFB242930E

[68] For the French Data Protection Authority (CNIL), ISO 27701 as a global standard, is not GDPR specific, nor does it constitute, as such, a GDPR certification instrument. "However, it represents the state of the art in terms of privacy protection and will allow organizations adopting it to increase their maturity and demonstrate an active approach to personal data protection.", 2 April 2020, https://www.cnil.fr/en/iso-27701-international-standard-addressing-personal-data-protection

[69] See H. Schepel, *The Constitution of Private Governance*, Oxford, Hart, 2005.

[70] See e.g., Communication from the Commission to the Council, the European Parliament and the European Economic and Social Committee, "Integration of Environmental Aspects into European Standardisation", COM/2004/130 final, 25 February 2004.

[71] On that issue, see the European standardisation strategy to reinforce public interest (i.e. based on EU policies) in standards: Communication from the Commission, "A Strategic Vision for European standards: Moving Forward to Enhance and Accelerate the Sustainable Growth of the European economy by 2020", COM/2011/0311 final, esp. point 10 on "using standards to address key societal challenges" such as consumer protection, climate change, accessibility, civil protection, personal data and individuals' privacy. Add. Communication from the Commission, "An EU Strategy on Standardisation Setting global standards in support of a resilient, green and digital EU single market", COM/2022/31 final, 2 February 2022, mentioning inter alia the development of standards "used to show compliance with rules imposed in the interest of EU citizens", p. 4.

Secondly, with regard to harmonised standards that are part of EU law, the valid criticisms of their current regime should not prevent them from being linked to fundamental rights considerations in sectors where there are strong adverse impacts on the said rights.[76] Advanced technologies, including AI, carry risks of a constitutional nature, in the sense that certain misuses of AI can undermine and destabilise Western democracies and the rule of law, including by invading the private sphere of European citizens.[77] On the contrary, integrating fundamental rights concerns into certain standards could help to give them an "ethical" dimension[78] and, at the same time, push for further reform of the European

---

[72] Regulation 1025/2012 on standardisation, *op. cit.* It imposes transparency obligations on standards bodies (national and European) in their work programmes and in the development of standards, in dialogue with the Commission. It provides that ESOs facilitate the participation of stakeholders outside industry, in particular SMEs, consumer, environmental and workers' associations as well as public authorities. See, in particular Articles 3 to 7.

[73] In practice, however, these stakeholders remain under-represented due to a lack of resources and expertise. In that respect, some see multi-stakeholder participation as an illusion. It is true that it requires a very large investment, just like the parliamentary law-making process. See K. Jakobs, R. Procter and R. Williams, "User participation in standards setting –The panacea?", *StandardView*, Vol. 6, no. 2, pp. 85-89, 1998; S. Moon and H. Lee, "The Primary Actors of Technology Standardization in the Manufacturing Industry", in *IEEE Access*, Vol. 9, pp. 101886-101901, 2021, doi: 10.1109/ACCESS.2021.3097800.

[74] Regulation (EU) 2022/2480 of the European Parliament and of the Council of 14 December 2022 amending Regulation (EU) No 1025/2012 as regards decisions of European standardisation organisations concerning European standards and European standardisation deliverables, OJ L 323, 19 December, 2022, p. 1-3.

[75] See Communication of the European Commission titled "An EU Strategy on Standardisation - Setting global standards in support of a resilient, green and digital EU single market", COM(2022)31, 1 February 2022, p. 5 (*the emphasis is mine*).

[76] See J-S Gordon and V. Fomin, "Ethics and Standardization", *in* (Ed.) Kai Jakobs, *Corporate Standardisation Management and Innovation*, IGI Global, Hershay, 2019, p. 177-192, §4.2.

[77] See H. Micklitz, O. Pollicino, A. Reichman, A. Simoncini, G. Sartor, and G. De Gregorio (Eds.), *Constitutional Challenges in the Algorithmic Society*, 2021, Cambridge University Press.

[78] It perfectly echoes the voluntary development of ethical guidelines by some tech companies. See for example "Atos blueprint for Ethics by design", described in *Global Opinion Paper on Digital Visions: Ethics, 2020*. See also the Ethical guidelines of IBM and Microsoft, referred by W. Barfield and U. Pagallo, *in*

standardisation process. Such a movement is perfectly in line with corporate social responsibility (hereinafter "CSR") and the duty of care that has developed in recent years on a global and local scale.[79] It is about holding multinational companies accountable for the human rights and environmental damages they cause through their supply chain around the world and in particular in the Global South. This is achieved through obligations of due diligence on their part and through a litigation framework based on civil liability regime. It also gave rise to an international standard, namely ISO 26000:2010 "Guidance on social responsibility". The latter provides guidance – but no requirements – for "assessing an organisation's commitment to sustainability and its overall performance",[80] including the respect for society and the environment based on fundamental rights.[81] The United Nations has also proposed a transposition of CSR regulatory framework to the digital ecosystem for tech multinational companies.[82] The EU AI Act thus seems to build on this work and reflect its rationale.

Finally, (ensuring) accountability of economic operators should encourage the integration of fundamental rights into standards on an *ad hoc* basis. Indeed, the respect of fundamental rights in the field of AI can only benefit from such a complementary approach, on the one hand, *ex ante* and bottom-up (coming from market actors) and, on the other hand, *ex post* and top-down (coming from hard law, as illustrated by the forthcoming AI Act).[83] Based on the NLF, compliance with harmonised standards constitutes for the producer or provider "a possible technical means to comply with [EU law]".[84] In the case of AI providers, they will be solely responsible for the risk assessment (provided for in the proposed regulation and based, inter alia, on a fundamental rights approach, as explained above) of their AI systems. And it is only on this basis that they will be able to determine the essential requirements with which they have to comply and how to comply with them, in the context of their organisation but also vis-à-vis individuals and the society. Hence, it could be very useful for them to have clear guidelines within harmonised standards on how to assess the fundamental rights dimension and implication of AI systems. Such standards would make the holistic and human-centred regulatory approach to AI more effective.[85] This

---

*Advanced introduction to AI*, Edward Elgar, 2021. Cf. J-S Gordon and V. Fomin, "Ethics and Standardization", *op. cit.*

[79] See the UN Guiding Principles on Business and Human Rights of 2011; OECD Guidelines for Multinational Enterprises (2000 and updated in 2011) and OECD Due Diligence Guidance for Responsible Business Conduct (2018). Cf. For a typology, S. Cossart, J. Chaplier, & T. Beau De Lomenie, The French Law on Duty of Care: A Historic Step Towards Making Globalization Work for All, *Business and Human Rights Journal*, 2017, 2 p. 317-323.

[80] https://www.iso.org/iso-26000-social-responsibility.html

[81] One of the principles of social responsibility developed by ISO 26000 is the respect of human rights (esp. §4.8 of the said standard).

[82] See Human Rights Council Resolution 47/23 on "New and emerging digital technologies and human rights", 16 July 2021.

[83] In practice, the legal consequences of this accountability will greatly depend on the effectiveness of the enforcement framework of the European regulation of AI systems – including harmonised standards – in terms of public enforcement. This does not mean, however, that the tech industry and AI practitioners should not have an *ex-ante* responsibility in the governance of AI technologies.

[84] "Blue Guide" on the implementation of the EU Product Regulation 2022, 2022/C 247/01, OJ C 247, 29.6.2022, p. 1-152, esp. §5.

[85] As proposed by M. Ebers, the European Commission could also "reconsider its approach" on standardisation (based on the NLF) and "establish legally binding obligations regarding the essential requirements for high-risk AI systems, such as what types of biases are prohibited; how algorithmic biases should be mitigated; and what type and degree of transparency AI systems should have, to name few", *in* "Standardizing AI – The Case of the European Commission's Proposal for an Artificial Intelligence Act", *op. cit.* (concluding remarks). By comparison it has been reported in the context of the EU AI Act negotiations

consideration is particularly relevant within the EU where the European institutions openly support the idea of trustworthy AI, enshrined in fundamental rights of citizens.[86]

Given the crucial role of fundamental rights in the governance of high-risk AI systems, including standardisation, the drafting of European standards "translating" fundamental rights considerations could be the way forward.[87]

## 3.2 First steps for integrating fundamental rights considerations into AI standards

### 3.2.1 Building on the international acquis on fundamental rights considerations in standardisation

Despite the legitimacy concerns towards the European standards-setting process, it should be pointed out that the inclusion of fundamental rights considerations in standards is not a new issue worldwide. The EU and ESOs surely have lessons to take from external experiences and research in this field. Some international organisations, both public and private, have already demonstrated great interest in the interplay between Information & Communication Technology (ICT) and advanced digital technologies (such as AI) standardisation and fundamental rights.

This is first the case of intergovernmental human rights organisations such as the United Nations and its Human Rights Council. The High Commissioner for Human Rights is currently leading a reflection on the relationship between fundamental rights and standardisation in the field of emerging digital technologies[88]. It builds on the 2021 Human Rights Council Resolution 47/23 on "New and emerging digital technologies and human rights".[89] The resolution calls for Member States "to place human rights at the centre of regulatory frameworks and legislation on the development and use of digital technologies".[90] Two potential ways to achieve this are mentioned: first, the application of the Guiding Principles on Business and Human Rights to the activities of technology companies and, second, technical standard-setting processes integrating a focus on fundamental rights considerations.

A second illustration may be found among standardisation bodies. Certain ISO standards already mentioned above directly address fundamental rights in relation to personal data protection and corporate social responsibility. In AI domain, ISO/IEC Joint Technical Committee 1/SC 42[91] released interesting technical reports (i.e., informational only)

---

that some Members of the European Parliament want the Commission "to issue common specifications on requirements for high-risk systems related to protecting fundamental rights. These specifications would be repealed once included in technical standards", *in* L. Bertuzzi, "AI Act: MEPs extend ban on social scoring, reduce AI Office role", 1 March 2023, *Euractiv* (online).

[86] On the "European approach to AI", see the dedicated webpage of the European Commission: https://digital-strategy.ec.europa.eu/en/policies/european-approach-artificial-intelligence

[87] Notably, it has been reported in the context of the EU AI Act negotiations that some Members of the European Parliament want the Commission "to issue common specifications on requirements for high-risk systems related to protecting fundamental rights. These specifications would be repealed once included in technical standards", *in* L. Bertuzzi, "AI Act: MEPs extend ban on social scoring, reduce AI Office role", 1 March 2023, *Euractiv*, available at: https://www.euractiv.com/section/artificial-intelligence/news/ai-act-meps-extend-ban-on-social-scoring-reduce-ai-office-role/?utm_source=substack&utm_medium=email

[88] See OHCHR consultation on human rights and technical standard-setting processes for new and emerging digital technologies, 15 February 2023.

[89] Human Rights Council Resolution 47/23, *op. cit.*

[90] Point 3 of the HRC Resolution 47/23, *op. cit.*

[91] https://www.iso.org/committee/6794475.html

interrelated with fundamental rights consideration, on "ethical and societal concerns posed by AI",[92] "trustworthiness"[93] and "bias".[94] As regards proper standards, works are ongoing on different aspects of AI, such as on management system.[95] In this context, assessing risk of AI could lead to different actions such as AI system impact assessment, including impacts on fundamental rights; but it seems rather unlikely that fundamental rights considerations will ultimately be given a prominent place. In addition, IEEE, as a large technical professional organisation dedicated to advancing technology, has long worked on the integration of ethical principles, including fundamental rights, into standardisation. In particular, the organisation has developed an ethical certification programme for autonomous and intelligent systems[96] that integrates a fundamental rights impact assessment into an ethical assessment of the systems.

To these initiatives could be added various national,[97] academic[98] and NGOs[99] proposals that develop compliance assessment frameworks for AI systems based on fundamental rights. Even if those frameworks are not per se meant to be used as standards, they could be a relevant source of inspiration for European standards-setters in AI and be integrated into standard deliverables.[100]

All this acquis is essential. It demonstrates, and thus confirms, the major challenge of respecting fundamental rights in the context of increased development and deployment of AI systems. The higher autonomy of these systems combined with the black-box phenomenon entail greater risk factors for individuals and the society. Risk management is key in standardisation.[101] The risks of infringement of fundamental rights must therefore be taken into account by stakeholders when drafting AI standards. Risk mapping in AI, for example, is certainly primarily technical. But it should also give rise to a dialogue with and translation into the conceptual sphere of fundamental rights.[102] In this context, the EU

---

[92] ISO/IEC TR 24368:2022, Information technology — Artificial intelligence — Overview of ethical and societal concerns.

[93] ISO/IEC TR 24028:2020, Information technology — Artificial intelligence — Overview of trustworthiness in artificial intelligence.

[94] ISO/IEC TR 24027:2021, Information technology — Artificial intelligence (AI) — Bias in AI systems and AI aided decision making.

[95] ISO/IEC DIS 42001, Information technology — Artificial intelligence — Management system.

[96] IEEE CertifAIEd™. All information available at https://engagestandards.ieee.org/ieeecertifaied.html

[97] See for instance the "Fundamental Rights and Algorithm Impact Assessment, named (FRAIA)" from the Dutch Government, *op. cit.* Cf. FRA Report, *Getting the future right: Artificial Intelligence and fundamental rights*, 14 December 2020.

[98] See for instance V. Gautrais and N. Aubin, Modèle d'évaluation des facteurs relatifs à la circulation des données, Instrument de protection de la vie privée et des droits et libertés dans le développement et l'usage de l'intelligence artificielle, Mars 2022, version 0.1, Available at: https://chairelrwilson.openum.ca/files/sites/36/2022/03/Modele_IA_Version_0.1.pdf ; H. Janssen, M. Seng Ah Leen J. Singh, Practical fundamental rights impact assessments, *International Journal of Law and Information Technology*, 2022, 30, p. 200-232.

[99] See for instance B. Nonnecke, P. Dawson, Human Rights Impact Assessment for AI: Analysis and Recommendations, October 2022, *AcessNow*.

[100] See *infra* 3.2.2.

[101] See in particular ISO 31000.

[102] As explained above, it is not for standards to set the level of risk of "acceptable" (or not) infringements of fundamental rights. This balance is a matter for the law, as shown by the "risk pyramid" in the AI Act which is set by the Union legislator. It is clear, however, that this is a major point of attention. See, expressing this fear vis-à-vis standards on AI if considering fundamental rights, M. Gornet, and W. Maxwell, "Intelligence artificielle: norms techniques et droits fondamentaux, un mélange risqué", 28 September 2022,

should develop a robust AI standardisation strategy geared towards this societal dimension of respecting the fundamental rights of Europeans.

### 3.2.2 Integrating fundamental rights considerations in the EU strategy on AI standardisation

Back in 2020 when the European Commission published its White Paper on AI,[103] CEN and CENELEC set up a working group on AI standardisation.[104] The experts prepared a "European roadmap for AI standardisation", aiming to support the European AI industry, while mitigating the risks for European citizens.[105] The document provides a general overview of AI standardisation activities at European and global level (i.e. IEEE, ETSI, ISO/IEC, ITU-T and CEN-CENELEC), and then identifies a series of areas on which European AI standardisation should focus, including stakeholder responsibility, data for AI, safety and privacy, ethics, engineering and safety of AI systems.[106] While some of these topics are already the subject of standardisation work, in particular within ISO, the expert's group also identifies "gaps" between the needs and expectations for standardisation in the EU and the current work in non-European standardisation bodies.[107] In particular, harmonised standards should specifically reflect "fundamental European values and human rights […]".[108] This statement is essential and perfectly consistent with the "humanist" dimension of the AI Act – placing respect for fundamental rights at the top of public interests to be safeguarded –, also reflected in the draft Standardisation Request. However, while giving an important indication on the potential orientation and dimension of future harmonised standards, it remains to be specified how this "Europeanisation", including fundamental rights considerations, of AI standards should (and will) be achieved. This is certainly one of the major challenges of future standardisation work, considering in particular the difficulty of standardising such evolving and complex technologies.[109]

CEN and CENELEC have created, in the second half of 2021, a permanent structure to conduct AI standardisation activities, in the form of a Joint Technical Committee (JTC), currently JTC 21. It is the point of contact for the European Commission as well as for the other standardisation bodies active in Europe in the field of AI standardisation. It should in principle provide the AI normative deliverables on the basis of the AI Act by 31 January 2025 at the latest.[110] It will have to formally accept the request for standardisation that will soon be notified to it by the Commission and to prepare a "work programme" indicating the standards to be drafted, the responsible technical bodies and a timetable for the

---

*The Conversation*, available at: https://theconversation.com/intelligence-artificielle-normes-techniques-et-droits-fondamentaux-un-melange-risque-189587

[103] COM/2020/65 final.

[104] CEN-CENELEC response to the EU White Paper in AI, June 2020.

https://www.cencenelec.eu/media/CEN-CENELEC/Areas%20of%20Work/Position%20Paper/cen-clc_ai_fg_white-paper-response_final-version_june-2020.pdf

[105] CEN-CENELEC Focus Group Report: Road Map in Artificial Intelligence (AI).

[106] Cf. S. Nativi and S. De Nigris, *AI Standardisation Landscape: state of play and link to the EC proposal for an AI regulatory framework*, EUR 30772 EN, Publications Office of the European Union, Luxembourg, 2021.

[107] *Ibid*.

[108] *Ibid*.

[109] See H. Pouget, "The EU's AI Act Is Barreling Toward AI Standards That Do Not Exist", 12 January 2023, LAWFARE, available at: https://www.lawfareblog.com/eus-ai-act-barreling-toward-ai-standards-do-not-exist

[110] Annex 1, draft standardisation request, *op. cit.*

execution of the requested standardisation activities.[111] In practice, several expert sub-groups are already working on operational aspects (e.g. risk and compliance), engineering (e.g. natural language processing, governance and data quality) and societal aspects (e.g. trustworthy AI characteristics, AI-enhanced nudges, sustainable AI).[112] It is probably mainly in this last sub-group that the issue of fundamental rights would be addressed. But it should also be considered at the strategic level and thus across all working groups.[113]

From now on, the central question relates to the methodology to be followed for a constructive dialogue on fundamental rights between all stakeholders in the standardisation of AI systems. The vast majority of these stakeholders represent the AI industry and have a technical background in both standardisation and AI engineering. Fundamental rights are usually not part of their vocabulary or their concern. This is fully understandable. However, it is clear that European public authorities intend to bring fundamental rights considerations to the forefront in AI standardisation, following societal and global concerns on this issue. Even if this may be open to valid criticism,[114] it is necessary to think about how to give concrete expression to this orientation within the ESOs. In our view, it is a work of dialogue and translation, supported by an effective regulatory framework.

Through *dialogue*, the objective would be to raise awareness and accountability among all stakeholders of the fundamental rights implications in the development and use of AI systems. The topic is very sensitive. On one side, it should be made clear that standards are not intended as such to implement fundamental rights. But on the other, standards should, on a case-by-case basis, take into account fundamental rights' sphere of protection in order to align AI technologies with it.[115] This first implies that NGOs, independent researchers and academics active in the field of advanced digital technologies, including societal and fundamental rights issues, take part in an open discussion with ESOs, including well-established players in standardisation (e.g. in the form of a series of workshops) on their role in European AI standardisation. This is already the case for the "Annex III organisations" under the EU Regulation on standardisation,[116] but this movement of inclusiveness must be reinforced. Concrete proposals have recently been put forward by those organisations (beyond the sole AI standardisation context)[117] ; they should be implemented by the ESOs in good time, under the supervision of the European

---

[111] Article 2, draft standardisation request, *op. cit.*

[112] Se presentation of S. Hallensleben, Chair CEN-CENELEC 21, 19 December 2022.
https://hsbooster.eu/sites/default/files/20221219_HSBoosterWebinar_SHallens.pdf (slide 4).

[113] As explained above (§2.2.2), based on the Standardisation Request, JTC 21 will have to justify to the European Commission the efforts undertaken to ensure, on the one hand, the multi-stakeholder participation of the AI ecosystem stakeholders, in particular civil society organisations, and, on the other hand, a particular expertise in the field of fundamental rights.

[114] See *supra* §3.1.1.

[115] Cf. C. Caeiro, K. Jones, E. Taylor, Technical Standards and Human Rights: The Case of New IP, in Ch. Sabatini (ed), *Reclaiming Human Rights in a Changing World order*, Brookings Institution Press, 2022, p. 185 et seq., Part. 3.

[116] See Annex III on "European stakeholder organizations eligible for union financing" in the field of standardization under EU Regulation 1025/2012 on standardization (i.e. ETUC, ANEC, ECOS and SBS)

[117] CEN and CENECLEC's governance review in support of inclusiveness. Proposal of the Annex III organizations, December 2022, available at https://www.etuc.org/sites/default/files/page/file/2022-12/Annex%20III%20proposals%20for%20CEN-CENELEC%20governance%20review_Dec%202022%20-%20Published%20version.pdf

Commission.[118] In other words, hard law should support voluntary normative process when fundamental rights are at stake. Secondly, vis-à-vis the industry representatives, there is certainly a need for simple practical guidelines setting out the European framework of fundamental rights and their interrelation with AI systems. There is already a lot of initiatives and proposals on the field that could facilitate this drafting.[119]

As far as *translation* is concerned, the objective would be to make fundamental rights considerations operational in the drafting of AI standards when relevant, for instance as regards actions to address risks such as AI systems impact assessment. This work should be greatly facilitated by the *dialogue* stage and the practical guidelines mentioned above. Here again it is the EU lawmakers – and alongside them the European Commission, guardian of EU law – that must set the "roadmap" for the fundamental rights considerations, as well as ensure a follow-up and monitoring of effectiveness. In that respect, the list of requirements laid down in the (draft) Standardisation Request and that echoes the AI Act should set out the contours and scope of fundamental rights considerations into the standardisation work. When relevant, normative deliverables could be illustrated and complemented by fundamental rights considerations based on EU law acquis and AI use cases. Based on Recital 9 of the draft Standardisation Request, it may also be justified to investigate in the relevance and adequacy of developing an "additional standard" within JTC 21, focusing on fundamental rights impact assessment for (high-risk) AI systems. Such a standard could then be combined with other AI standards, including risk mapping, risk management, trustworthy AI and conformity assessment. Its value would be to address the issue of fundamental rights implications in a more systematic and detailed way, complementing the more classical developments on "ethical and societal impacts" of other standards.

## 4 Conclusions

There is still a long way to go before the AI Act is adopted and implemented. However, it is already clear that harmonised standards, whether newly developed in Europe within ESOs or take up from international standardisation organisations deliverables, will play a key role. AI system providers, but also other stakeholders in the AI value chain, will have to be able to rely on the detailed, manageable and practical tools and methodologies provided by standards to make their regulatory compliance with the AI Act operational, in the service of a trusted European digital ecosystem. In this context, fundamental rights considerations should play an essential role. They are the direct translation of the highest EU public interests that harmonised standards should uphold in the European digital market, as well as the bulwark against unreasonable risks for the society in the face of the rapid progress of AI technologies.

In that context, some recommendations dealing with fundamental rights considerations in AI standards-setting process can be formulated on the basis of this contribution.

**1:** The inclusion of human rights considerations in the standardisation of AI must be well understood. This is not a "privatisation" to standardisation bodies (and their members) of human rights advocacy in the field of AI. ESOs are not being asked to develop standards for, or dealing *per se* with the protection of fundamental rights in AI systems. Rather, human rights considerations in standard-setting process are a way of raising awareness and accountability as regards the major public and societal challenges posed by AI. AI systems deployments and uses pose new and far-reaching risks to individuals and society. Standardisation has its role to play, as does hard law. Based on this cross-cutting normative perspective, the question of human protection (e.g., human dignity and freedom) and societal protection (e.g., democracy and the rule of law) in the AI ecosystem is crucial. Therefore, ESOs should develop standards that can protect AI subjects from the risk of fundamental rights violations from AI systems.

---

[118] On the same opinion and for other complementary proposals, see Ch. Galvagna, Discussion paper: Inclusive AI governance, Civil society participation in standards development, Ada Lovelace Institute, 30 March 2023.

[119] See the references cited above in §3.2.1. Cf. in the field of Internet standardization, the work of the Human Rights Protocol Considerations Research Group (HRPC) within the Internet Research Task Force (parallel organization of the Internet Engineering Task Force (IETF).

**2:** In view of the legitimacy crisis of standardisation, at least in Europe, the central role given to standards in the field of AI should serve as a use case for the evolution of the institutional framework of standardisation. A "constitutional" turning point is needed. In particular, for harmonised standards whose mere reference is published in the EU Official Journal, a paradigm shift should take place: they should be fully accessible (i.e., *in extenso*) to all citizens. Such a request of free access to standards is made, in a more moderate way, by the "Annex III organisations" under the EU Regulation on standardisation to ensure their effective involvement in the co-construction of standards, whether or not they are future harmonised standards. This proposal should apply beyond these organisations and, at least, benefit all academic researchers. It is indeed an indispensable means for academic research to invest in the critical analysis of standards, in particular when they have huge societal implications, as in the field of AI. In addition, for harmonised standards that involve fundamental rights considerations, an *ad hoc* drafting and adoption scheme with mandatory participation of (at least) European legislative representatives (in particular Members of the European Parliament) in the standardisation work should be created.[120] Less ambitiously, Member States are also expected to implement Article 7 of the EU Regulation on standardisation according to which they should encourage the participation of public authorities, including market surveillance authorities, in standardisation activities. This includes inter alia data protection authorities, whose expertise in AI is growing.

**3:** With regard to the ongoing preparation of harmonised standards for AI systems on the basis of the AI Act, there is an urgent need for awareness and accountability at two levels. The first level is aimed at representatives of civil society, who should be part of the standardisation work on AI, whatever form this integration takes. (e.g., open discussion, workshops series, critical reviews of ongoing work items, etc., within JCT 21). The second is aimed at established standardisation actors, in particular representatives of large tech organisations active in the AI field. The latter should undertake a dialogue between technical – state of the art – and societal – including fundamental rights considerations – aspects of AI, and its implementation into standards. Several tools and actions can be imagined and implemented to this end (e.g., preparation of guidelines on EU fundamental rights and their implications in AI use cases, regular critical reviews of ongoing work items by non-technical experts with specialisation in law, sociology, ethics, etc., within JCT 21). Their effectiveness should be supported by hard law requirements, which would require – from a general perspective – a new revision of the EU Regulation on standardisation, more ambitious than the last one. We could also imagine a compliance monitoring of the

---

[120] In the context of the current negotiations on the AI Act, the European Parliament is proposing that the technical specifications (which fall within the competence of the European Commission) should deal with fundamental rights concerns (and not European harmonised standards). See Article 41(1) (b) of the Consolidated MCOLIBE AI Act, 14 May 2013 (Draft Compromise Amendments on the Draft Report, 9 May 2023).

European standards, in particular as regards the protection of EU public interests, including fundamental rights, by the (future) EU AI regulatory authorities, based on the AI governance scheme proposed by the AI Act.